\documentclass[a4paper,fleqn]{cas-dc}

\usepackage[numbers]{natbib}
\usepackage{algorithm}
\usepackage{algorithmicx}
\usepackage{algpseudocode}
\usepackage{subfigure}
\usepackage{amsmath}
\DeclareMathOperator*{\argmax}{arg\,max}

\usepackage{todonotes}
\usepackage{xpatch}

\makeatletter
\xpatchcmd{\@todo}{\setkeys{todonotes}{#1}}{\setkeys{todonotes}{inline,backgroundcolor=yellow,#1}}{}{}
\makeatother

\def\name{TransMUSE}
\def\tsc#1{\csdef{#1}{\textsc{\lowercase{#1}}\xspace}}
\tsc{WGM}
\tsc{QE}

\begin{document}
\let\WriteBookmarks\relax
\def\floatpagepagefraction{1}
\def\textpagefraction{.001}

\shorttitle{TransMUSE}    

\shortauthors{L. Xu et al.}  

\title [mode = title]{TransMUSE: Transferable Traffic Prediction in MUlti-Service Edge Networks}  



%

\author[1,2,3]{Luyang Xu}
\ead{xuluyang@cnic.cn}
\cormark[1]
\credit{Conceptualization, Methodology, Formal Analysis, Experiments, Writing - original draft, review and editing}

\author[2]{Haoyu Liu}
\ead{haoyu.liu@ed.ac.uk}
\credit{Methodology, Writing - review}

\author[1]{Junping Song}
\ead{songjunping@cnic.cn}
\credit{Data Cooperation and Analysis, Writing - review}

\author[4]{Rui Li}
\ead{rui.li@samsung.com}
\credit{Conceptualization, Writing - review}

\author[5]{Yahui Hu}
\ead{huyahui@cumtb.edu.cn}
\credit{Conceptualization}

\author[1]{Xu Zhou}
\ead{zhouxu@cnic.cn}
\cormark[1]
\credit{Conceptualization, Resources, Supervision, Project administration}

\author[2]{Paul Patras}
\ead{Paul.Patras@ed.ac.uk}
\credit{Conceptualization, Writing - review and edit, Supervision, Project administration}

\affiliation[1]{organization={Computer Network Information Center, Chinese Academy of Sciences},
            addressline={Building No.2, 4, Zhongguancun Nansijie, Haidian District}, 
            city={Beijing},
            postcode={100190}, 
            state={Beijing},
            country={China}}

\affiliation[2]{organization={School of Informatics, The University of Edinburgh},
            addressline={10 Crichton Street}, 
            city={Edinburgh},
            postcode={EH8 9AB}, 
            country={United Kingdom}}

\affiliation[3]{organization={University of Chinese Academy of Sciences},
            addressline={No.19(A) Yuquan Road, Shijingshan District}, 
            city={Beijing},
            postcode={100049}, 
            country={China}}
            
\affiliation[4]{organization={Samsung AI Center},
            addressline={50 Station Road}, 
            city={Cambridge},
            postcode={CB1 2JH}, 
            country={United Kingdom}}

\affiliation[5]{organization={China university of mining and technology-Beijing},
                addressline={Ding No.11 Xueyuan Road},
                city={Beijing},
                postcode={100083},
                country={China}}
\cortext[1]{Corresponding author}



\begin{abstract}
The Covid-19 pandemic has forced the workforce to switch to working from home, which has put significant burdens on the management of broadband networks and called for intelligent service-by-service resource optimization at the network edge. In this context, network traffic prediction is crucial for operators to provide reliable connectivity across large geographic regions. Although recent advances in neural network design have demonstrated potential to effectively tackle forecasting, in this work we reveal based on real-world measurements that network traffic across different regions differs widely. As a result, models trained on historical traffic data observed in one region can hardly serve in making accurate predictions in other areas. Training bespoke models for different regions is tempting, but that approach bears significant measurement overhead, is computationally expensive, and does not scale. Therefore, in this paper we propose TransMUSE (Transferable Traffic Prediction in MUlti-Service Edge Networks), a novel deep learning framework that clusters similar services, groups edge-nodes into cohorts by traffic feature similarity, and employs a Transformer-based Multi-service Traffic Prediction Network (TMTPN), which can be directly transferred within a cohort without any customization. We demonstrate that TransMUSE exhibits imperceptible performance degradation in terms of mean absolute error (MAE) when forecasting traffic, compared with settings where a model is trained for each individual edge node. Moreover, our proposed TMTPN architecture outperforms the state-of-the-art, achieving up to 43.21\% lower MAE in the multi-service traffic prediction task. To the best of our knowledge, this is the first work that jointly employs model transfer and multi-service traffic prediction to reduce measurement overhead, while providing fine-grained accurate demand forecasts for edge services provisioning.
\end{abstract}




\maketitle


\section{Introduction}
Edge computing pushes computation and data storage closer to the user, thereby improving response times and saving communication bandwidth, while serving multiple applications simultaneously, e.g., video streaming, gaming, content delivery, etc. 
As people work increasingly more often remotely following the Covid-19 outbreak and require network support for different services, the edge computing paradigm is witnessing growing uptake. 

In order to optimise user experience and operational costs, infrastructure providers have been pursuing dynamic provisioning of network resources based on predictions of user demand \cite{fang2021spider}. Previous efforts in tackling network traffic prediction frequently exploit the ability of deep neural networks (DNNs) to learn complex patterns from historical data \cite{chang2018memory, huang2017study, zhang2018long, wang2018spatio,he2020graph, zhang2019multi}. However, existing solutions either require training one dedicated model for each geographic region and hence have limited transferability (which is of paramount importance in reducing computational costs and the environmental footprint of training DNNs) \cite{chang2018memory, huang2017study, zhang2018long, wang2018spatio}, or disregard essential correlations among services \cite{he2020graph, zhang2019multi}. 

\begin{figure}[t]
    \centering
    \includegraphics[width=\linewidth]{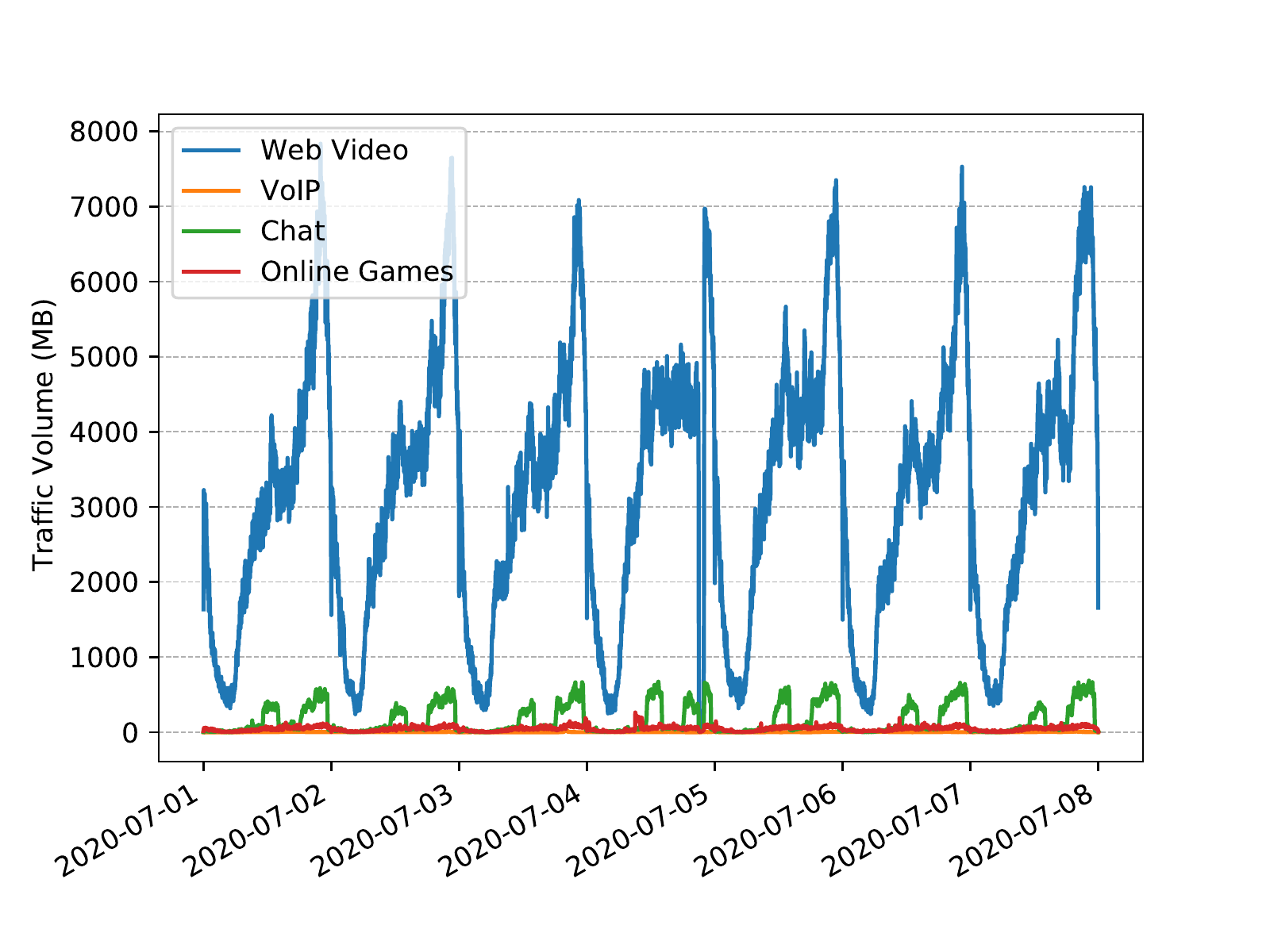}
    \caption{A snapshot of the volume of traffic consumed by four different services as observed at an edge node in a network deployment in Sichuan (China) over one week. 
    }
    \label{fig:service-example}
\end{figure}
In practical large-scale network deployments \emph{(i)} per-service patterns are often distinct within a region, as exemplified in Figure \ref{fig:service-example}, while \emph{(ii)} certain areas may exhibit similar characteristics that would allow for direct transfer of models among them, without the need of retraining. These key observations are confirmed by our analysis of a real-world network traffic dataset collected in a major city in Sichuan province, China, serving 2.6 million users, spanning 6.3 square kilometers, and comprising eight edge nodes. This motivates us to propose \name{}, a transferable traffic prediction framework in multi-service edge networks, which first groups edge-nodes according to per-service statistical features. Within each cohort, reference neural models are chosen and trained on data collected only in the region with the highest overall traffic consumption, which can be then transferred to other group members. As reference model, we put forward a Transformer-based~\cite{vaswani2017attention} Multi-service Traffic Prediction Network (TMTPN). Furthermore, we propose WK-means, a service clustering algorithm based on Wasserstein distance to categorize services according to their similarity. We train separately a TMTPN model for each service cluster to boost prediction performance at a regional level. Finally, the reference models are transferred to other regions directly, without adaptation. 

Our proposed model transfer framework, \name{}, provides a comprehensive and cost-effective solution for traffic prediction in multi-service edge networks. The key advantages of \name{} are as follows: \emph{(i)} it provides a model transfer approach among edge nodes to reduce measurement and computational overhead without compromising prediction accuracy -- compared with training a model individually on local data for each edge node, \name{} exhibits imperceptible performance degradation, with only 1.7\% and 0.26\% higher MAE and RMSE, respectively; \emph{(ii)} the proposed TMTPN takes service correlation into consideration to further reduce overhead and the energy that would have otherwise been required to maintain a separate prediction model for each service; our experiments demonstrate that TMTPN outperforms the state-of-the-art MTNet benchmark \cite{chang2018memory} on the multi-service traffic prediction task by 18.74\% and 18.49\%, in terms of MAE and respectively RMSE; \emph{(iii)} the WK-means service clustering tackles both model under-fitting and speed of convergence, improving the TMTPN prediction performance, as it attains 17.59\% and 27.89\% lower MAE and respectively RMSE, as compared to predicting without prior service clustering. To the best of our knowledge, \name{} is the first multi-service traffic forecasting solution for edge networks that leverages model transfer and service clustering to achieve high accuracy at a low measurement cost.

The rest of the paper is organised as follows. The multi-service prediction problem is formalised in Section~\ref{sec:problem}. The proposed \name{} framework is discussed in detail in Section ~\ref{sec:framework}. Section \ref{sec:experiments} provides exhaustive experimental results to demonstrate \name{}'s efficacy. Section \ref{sec:related-work} discusses the most relevant related work and Section \ref{sec:conclusions} concludes the paper.

\section{Problem Formulation}
\label{sec:problem}
Our aim is to address the challenges of handling spatial heterogeneity of service traffic in edge networks and reducing model training costs when forecasting future demands in edge networks, to support the effective management of their resources.

Formally, multi-service traffic forecasting seeks to maximize the probability that, given $T$ previous measurements of the traffic volume consumed by $K$ services, the predicted traffic consumption over $F$ future time steps is as close as possible to the ground truth. Denoting by $x_{t}^{k}$ the traffic volume of service $k$ at timestamp $t$ and $\mathcal{X}_{t}:=[x_{t}^{1}, ..., x_{t}^{K}]$ the snapshot of all $K$ services at time $t$, and considering a forecasting model that is parameterized by $\mathbf{\theta}$, the multi-service traffic forecasting problem is equivalent to:
\[
\argmax_{\theta} \; p_{\theta}\left(\mathcal{X}_{t+1}, ..., \mathcal{X}_{t+F} | \mathcal{X}_{t-T+1}, ..., \mathcal{X}_{t}\right).
\]
To solve this problem, we design a Transformer-based Multi-service Traffic Prediction Network (TMTPN) that captures temporal correlations among traffic time series via multi-head attention, then improve forecasting accuracy via service-clustering, as we detail next.

\begin{figure*}
    \centering
    \includegraphics[width=0.9\textwidth]{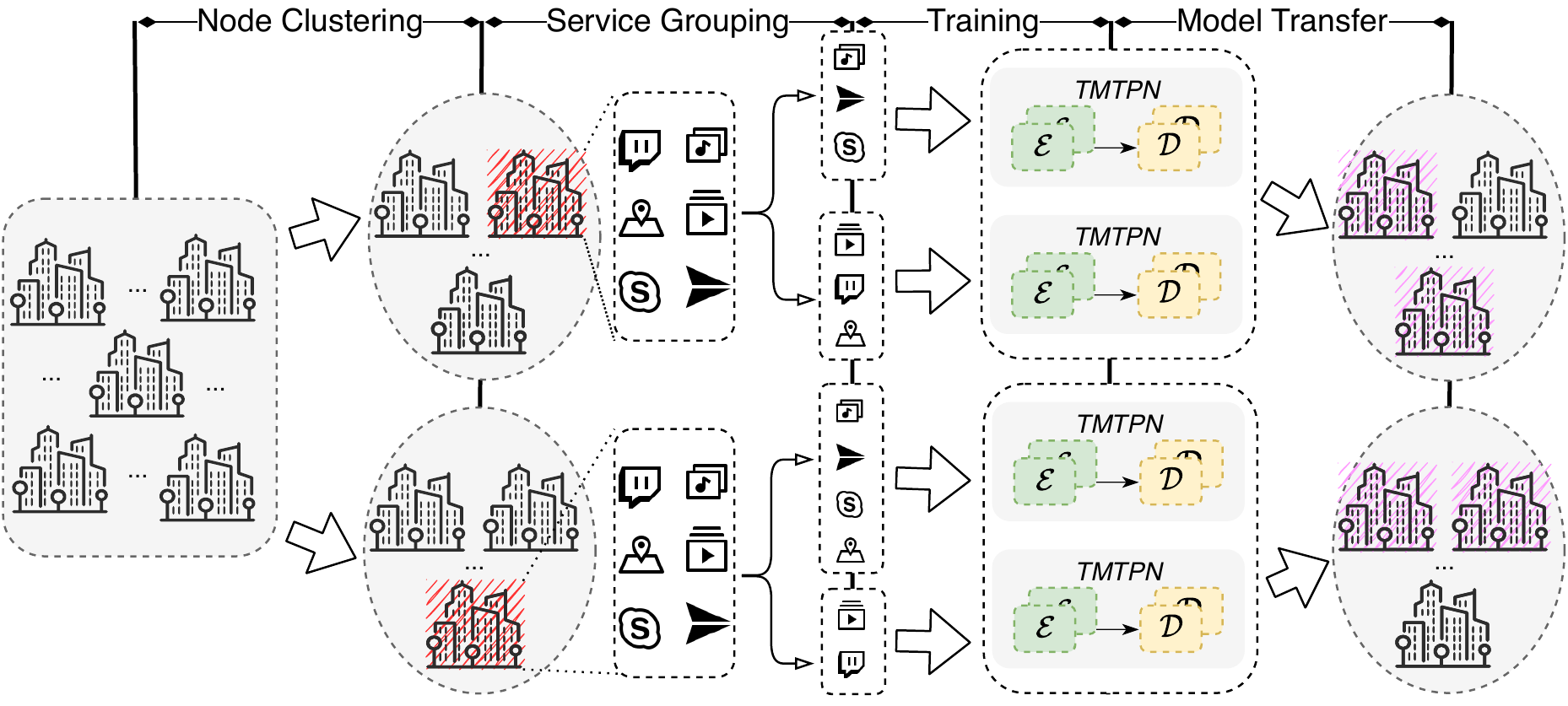}
    \caption{The proposed \name{} framework incorporates five stages: 1) Edge nodes are grouped into several \textit{node clusters}, within which model transfer is to be conducted; 2) In each cluster, the edge node with the largest traffic volume is selected as reference (highlighted with hashed patterns); 3) At each reference node, services are further partitioned into \textit{service clusters} by WK-Means; 4) One TMPTN is trained for each service cluster;  5) The models trained on \textit{reference nodes} are transferred to the recipients (highlighted on the right) within the corresponding node clusters.}
    \label{fig:\name{}}
\end{figure*}

\section{\name{} Framework}
\label{sec:framework}
We propose \name{}, a deep learning framework for accurate and cost-effective multi-service forecasting at the network edge. Figure \ref{fig:\name{}} gives an overview of the different components this framework entails and the relationship between them, namely:
\begin{enumerate}
    \item \textbf{Edge Node Clustering}: We cluster edge nodes by a set of service-level statistical features using the K-means algorithm, to determine the neural model transfer scope. 
    \item  \textbf{Reference Node Selection:} Within each scope, we select the node with the overall highest traffic consumption as the reference node; reference neural models for forecasting will be trained with data collected at such nodes.
    \item \textbf{Service Grouping:} As certain mobile services exhibit statistical similarities, we cluster services using a modified K-means algorithm based on Wasserstein distance, aiming to reduce the number of multi-service neural models to be employed for prediction.
    \item \textbf{Model Training:} At the level of each reference node, we train a dedicated TMTPN model for each service cluster, which will simultaneously predict the volume of traffic for all services within such clusters.
    \item \textbf{Model Transfer:} We transfer the trained reference models from each reference edge node to all other nodes within the corresponding clusters, where they will be applied for inference without further training.
\end{enumerate}
Next, we discuss in details the key stages of our \name{} framework.

\subsection{Edge Model Transfer}
In Multi-access Edge Computing (MEC) scenarios, it is often impractical to train a neural model at each individual edge node, as their computational power is limited and the operational costs and energy expenditure can become prohibitive to operators when deployment density increases. Edge model transfer aims to reduce the cost of measurement collection and model training, by confining these tasks to designated nodes and reusing  models trained there on other nodes, without further local tuning. Different from cloud--edge approaches where a central node maintains a global model refined through model updates resulting from local training (federated learning), edge model transfer only considers the model to be transferred among edge nodes without the need for a central cloud. This brings additional merits in terms of data privacy and communication overhead reduction, as the transfer process is confined within a limited scope. A model to be transferred is called a \textit{reference model}, the node where a reference model is trained is called a \textit{reference node}, and the edge nodes that adopt it are referred to as \textit{recipients}.

There are two key issues to address in the edge model transfer process. The first, is determining the scope of model transfer. Different edge nodes may observe distinct traffic patterns due to geographic dissimilarities in terms of mobile user demographics \cite{singh2019urban} or socioeconomic function (residential areas, business districts, shopping centers, etc.). Edge model transfer, therefore, should be applied across edge nodes (within a cluster) with similar traffic features. Secondly, choosing at which edge node to train a reference model to be transferred within the corresponding cluster will impact the inference accuracy. We put forward an edge model transfer strategy that deals with these two issues as follows:
\begin{itemize}
    \item \textbf{Determining Transfer Scope}: We use K-means clustering to group edge nodes according to four statistical features, i.e., \textit{mean}, \textit{standard deviation}, \textit{maximum} and \textit{minimum value} of traffic volume for each service over one month. With 20 services, this leads to vectors of shape $80 \times 1$ that represent an edge node.
    A model will only be transferred within the same cluster of edge nodes.
    \item \textbf{Reference Model Training}: Within each cluster, a reference model will be trained only with data from the edge node where the overall highest traffic consumption is observed. Reference models are then transferred to the recipients within the corresponding clusters. The results we present in Section~\ref{sec:experiments} confirm the generalization abilities of this approach.
\end{itemize}

At the level of a reference node, a set of neural models will be trained, each of which targets future traffic predictions for groups of services with similar characteristics, as we explain next.

\subsection{WK-means Service Clustering}
Traffic patterns and volumes may differ among services due to content popularity, number of service subscribers, service scope, etc. (see Figure~\ref{fig:service-example}), leading to high information entropy if observing all services together. Therefore, training a single model to predict the demand of all services may lead to under-fitting problems, because the model may need to learn highly convoluted patterns. To tackle this issue, we propose a service clustering algorithm based on the Wasserstein distance (WD) between per-service time-series data points, which we name \textit{WK-means}. This facilitates effective simultaneous predictions of future traffic volumes for services with similar sequential features.

There are two key factors to consider when measuring time-series similarity, namely, magnitude and `shape'. The former indicates how comparable the traffic volume of different services is; the latter indicates any similarities in terms of periodicity and short-term temporal patterns. The WD takes these two factor into consideration at the same time, which makes it particularly suitable for our grouping task. Originally the WD was proposed to measure the similarity between two probability distributions, and was recently employed in optimal transportation problems~\cite{shen2018wasserstein}:
\begin{equation}\label{eq_wasserstein}
W_{p}(\mathcal{P}, \mathcal{Q}) = \left(\inf_{\mu \in \Gamma(\mathcal{P}, \mathcal{Q})} \int \rho(x,y)^pd\mu(x,y)\right)^{1/p},
\end{equation}
where $\mathcal{P}$ and $\mathcal{Q}$ are two probability distributions in $\mathbb{R}^d$, and $\Gamma(\mathcal{P}, \mathcal{Q})$ is the set of all probability measures on $\mathbb{R}^d \times \mathbb{R}^d$ with marginals $\mathcal{P}$ and $\mathcal{Q}$. $\rho(x,y)^p$ is a measure of distance between $x \in \mathcal{P}$ and $y \in \mathcal{Q}$ (e.g., $p=2$ for Earth mover's distance). Intuitively, the WD represents the minimal distance for moving the mass of distribution $\mathcal{P}$ to exactly fit the mass of distribution $\mathcal{Q}$.

Unlike other distance metrics, such as Euclidean distance, Jensen–Shannon (JS) divergence or Kullback–Leibler (KL) divergence, WD has the following key advantages: 
(i) if the target distributions lie in low-dimensional manifolds or share disjoint support, which is not uncommon for high-dimensional data, WD offers a more informative measure (which is not the case for KL and JS divergence that return a constant value or infinity)
\cite{liu2019wasserstein, arjovsky2017wasserstein}; and
(ii) WD maintains the underlying geometry of the space \cite{cmu}, that is, it not only takes the quantitative value into consideration, but also pays attention to the similarity of distributions' shapes. In contrast, the Euclidean distance cannot quantify shape differences or capture the degree of changes between two times series \cite{dong2006research}.

Based on WD, we propose the WK-means service clustering algorithm, summarized by the pseudo-code in Algorithm \ref{alg1}. To generate $N$ clusters from $S$ services, WK-means initially sorts all the services by their volume and splits the sorted sequence at $\left[\frac{S}{N}, \frac{2S}{N}, ..., \frac{(N-1)S}{N}\right]$. That is, the sorted sequence is evenly divided into $N$ segments ({lines 2-6}). WK-means further chooses the service in the middle of each segment as cluster center ({lines 8-13}). Instead of using random initialization, this approach speeds up the convergence process. Then, the WD between each service and sub-cluster center is calculated, and each service is re-assigned to its nearest sub-cluster ({lines 14-24}). Finally, each sub-cluster center is updated ({line 25}) and the previous two steps are iterated until all sub-clusters convergence or the iteration epoch reaches a predefined limit ({lines 7-26}).

In a setting with multiple edge nodes, it is possible that WK-means may generate different service clustering results on different nodes. To comply with our model transfer strategy, we first conduct WK-means at the level of every edge-node and select the most frequent clustering pattern as the global service grouping. Such pattern is applied to all other nodes in our design. 
\begin{algorithm}
	\caption{WK-means Service Clustering} 
	\label{alg1} 
	\begin{algorithmic}[1]
		\Require $X$(list::service time series), $N$(int::cluster), $I$(int::max iteration) 
		\Ensure $L$(list::cluster results)
		\State $mean\_list = [\,]$, $centers\_list = [\,]$, $counter = 0$
		\For{$i=0 \to len(X)$}
		    \State $mean\_list.append(mean(X[i]))$
		\EndFor
		\State $mean\_list = sorted(mean\_list)$
		\State Initialize $X$ into $N$ clusters by $mean\_list[0:len(X):len(X)/N]$
		\Repeat
		\For{each cluster $C$}
            \If{$mean(X[i]) == median(C)$} 
                \State $C_{center}=X[i]$
            \EndIf     
		\State $centers\_list.append(C_{center})$
		\EndFor
		\For{$i=0 \to len(X)$}
		\State $min\_dist = 10000$
		\State $flag = 0$
		\For{$j=0 \to N$}
		\State $dist_{ij} = WD(X[i],$ $centers\_list[j])$
		\If{$dist_{ij} <= min\_dist$} \\
		\qquad \qquad \qquad {$min\_dist = dist_{ij}$ and $flag = j$}
		\EndIf
		\EndFor
		\State $L[i] = flag$ 
		\EndFor
		\State update each cluster \& $counter++$
		\Until $counter \geq I$ or $L$ is unchanged
	\end{algorithmic} 
\end{algorithm}

\begin{figure}[ht]
    \centering
    \includegraphics[width=0.9\columnwidth]{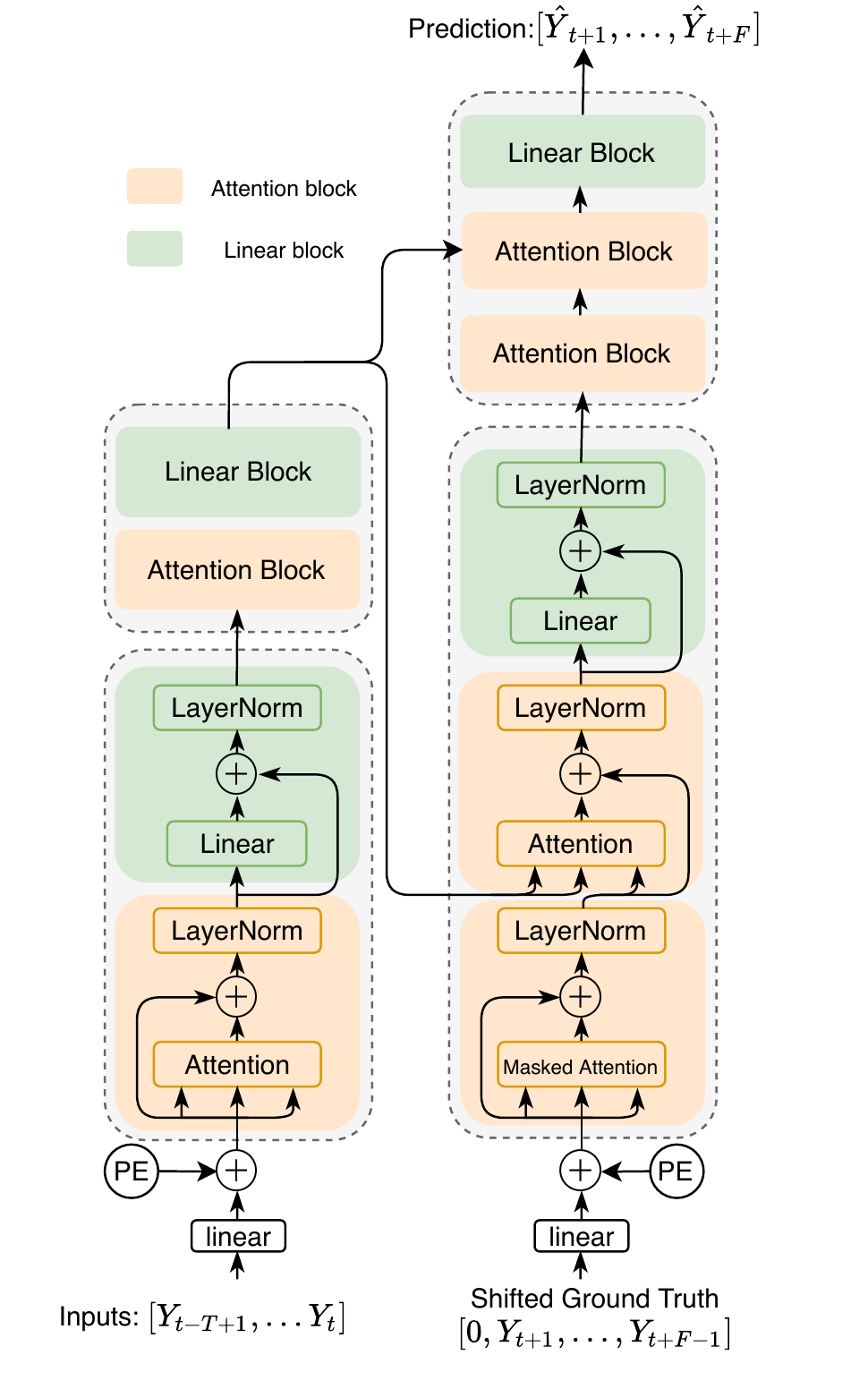}
    \caption{TMTPN architecture based on an Encoder-Decoder design. Historical traffic input combined with positional encoding (PE) is processed by the Encoder, which gives the output to each Decoder block. Within each, the core components are a multi-head Attention block and a Linear block.} 
    \label{fig:fig_TMTPN}
\end{figure} 

\subsection{TMTPN Model Design}
To perform multi-service traffic forecasting, we design Transformer-based Multi-service Traffic Prediction Networks (TMTPNs), each of which inherits from the canonical Transformer architecture and is dedicated to each individual service cluster. Transformers have shown remarkable performance in processing sequential data and have been adopted previously for natural language processing \cite{devlin2018bert}, computer vision \cite{pan20213d}, and vehicular traffic prediction \cite{xu2020spatial} tasks. 
Our TMTPN model is illustrated in Figure \ref{fig:fig_TMTPN} and follows an Encoder-Decoder paradigm, encompassing the following components:
\begin{itemize}
    \item \textbf{Multi-Head Attention:} 
    Multi-head attention consists of multiple scaled dot-product attention structures that capture temporal dependencies in long sequences. The attention block receives three inputs: $Q \in \mathbb{R}^{T \times d_{k}}$  (query), $ K\in \mathbb{R}^{T \times d_{k}} $ (key) and $ V\in \mathbb{R}^{T \times d_{k}} $ (value), in which $T$ represents the sequence length and $d_{k}$ is the embedding dimension of each item in the sequence. Attention is computed as:
    \begin{gather*}
        Attention(Q,K,V) = \text{softmax}\left(\frac{QK^{T}}{\sqrt{d_k}}\right)V.
    \end{gather*}
    $QK^{T}$ generates a $T \times T $ matrix of alignment scores, where each entry denotes the correlations between two instances in the sequence. The matrix is scaled and then multiplied by $V$ to generate the hidden representation of the input that incorporates attention information. Multi-head attention splits $Q$, $K$ and $V$ into multiple chunks, which are processed with independent attention blocks. The outputs of all the attention blocks are concatenated and projected back into hyperspace $\mathbb{R}^{d_{k}}$.
    
    \item \textbf{Encoder and Decoder Layers}: The encoder layers (Figure~\ref{fig:fig_TMTPN} left) contain a multi-head attention block and a linear block, each of which utilizes a skip connection and layer normalization to prevent over-fitting. For the encoder, only the input sequence is given to the multi-head attention block, i.e., \mbox{$X=Q=K=V$}, where self-attention is computed. The decoder (Figure~\ref{fig:fig_TMTPN} right) incorporates an extra attention block. Specifically, the first attention block in the decoder computes the self-attentional representations of the decoder input, and the second block takes the encoder output as the key $K \in \mathbb{R}^{T \times d_{k}}$ and the value $V \in \mathbb{R}^{T \times d_{k}}$, querying which historical inputs are important when making future predictions.  
    
    \item \textbf{Positional Encoding (PE):} Since transformers do not contain any sequential structure, timing features are not encoded in the network by default. Therefore, positional encoding is added to the input sequence, which reflects the relative position of each timestamp.
    PE is computed as:
    \begin{gather*}
        PE(pos,2i) = \sin(pos/10,000^{2i/d_{model}}), \\
        PE(pos, 2i+1) = \cos(pos/10,000^{2i/d_{model}}),
    \end{gather*} where $pos$ denotes the position index of the item in the sequence and $d_{model}$ is the dimension of the encoded position.
    
    \item \textbf{Parallel Decoding:} Traditional seq2seq models \cite{zhang2019multi} perform decoding in an auto-regressive manner during training. That is, decoding the $t^{th}$ element in a sequence relies on the hidden states passed from timestamp $t-1$ and the decoded \mbox{$(t-1)^\text{th}$} item, which are provided as the input. It is therefore impossible to decode all the items in parallel. Transformers overcome this problem during training by introducing the \textit{shifted decoder input} and \textit{look-ahead mask}. Assume that the ground truth to be provided to the decoder is $Y = [y_{1}, ..., y_{F}]$, then the input is the shifted-right ground truth $X = [0, y_{1}, ..., y_{F-1}]$. The \textit{look-ahead mask} ($M$) is introduced when computing the alignment scores as follows:
    \begin{gather*}
    A = \text{softmax}\left(\frac{QK^{T}}{\sqrt{d_k}}M\right),
    \end{gather*}
    where $X = Q = K$, and M is a $F \times F$ matrix with each entry above the diagonal equal to negative infinity, and below/on the diagonal equal to 0. The scaled matrix of alignment scores is masked with M, which yield a $F \times F$ lower triangular matrix, meaning that at a given timestamp $i$, there is no correlation ($A_{ij} = 0$) with the input from any future timestamp $j ~(j > i)$. By masking, the decoder can approximate the output at $F$ timestamps, $\hat{Y} = [\hat{y}_{1}, ..., \hat{y}_{F}]$, in parallel.
    This technique is only applied during training, while during testing the transformer decodes step-by-step, as seq2seq models.
\end{itemize}


Overall, the proposed TMTPN architecture has several merits: (i) it can be trained fast due as the look-ahead mask and the shifted decoder input that facilitate parallelization; (ii) it can process longer sequences than traditional seq2seq models; and (iii) it captures the most essential historical information that impacts most on prediction results, irrespective of the length of an input sequence, thanks to Position Encoding and Multi-Head Attention.

\section{Experiments}
\label{sec:experiments}

We implement \name{} and its TMTPN models, as well as a set of benchmark neural models in Tensorflow v2.3.0 using the cuDNN v7.6 and CUDA v10.1 libraries. To demonstrate the performance gains of our solution, we train and evaluate the neural models and experiment on a large-scale real-world wired network traffic dataset collected by a network operator in Sichuan Province, China. For this, we employ a high-performance computing cluster comprising 12 servers, each equipped with a 32-cores Intel E5-2620 CPU and running Red Hat Enterprise Linux, and accelerate the training process with multiple GPUs out of a pool of 96 Nvidia RT2080Ti units.

We conduct three sets of experiments to demonstrate (1) multi-service traffic prediction performance gains attained by our TMTPN models; (2) the benefit of employing service clustering with WK-means; and (3) forecasting performance with edge model transfer. 

\subsection{Dataset \& Pre-processing}
\begin{figure}
    \centering
    \includegraphics[width=\linewidth]{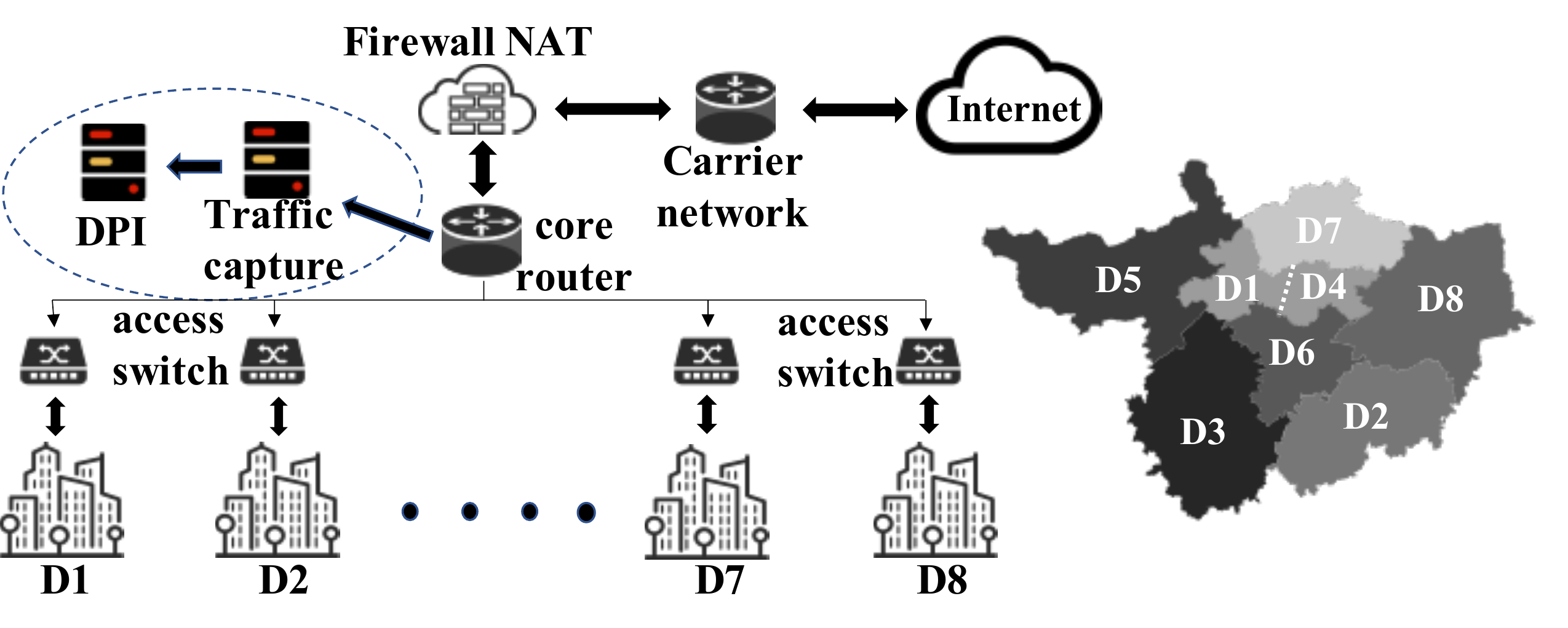}
    \caption{Network topology and district map of the target city. Traffic measurements collected via DPI and further processed at the core router.}
    \label{fig:fig_map}
\end{figure}

\begin{figure}[ht!]
    \centering
    \includegraphics[width=\linewidth]{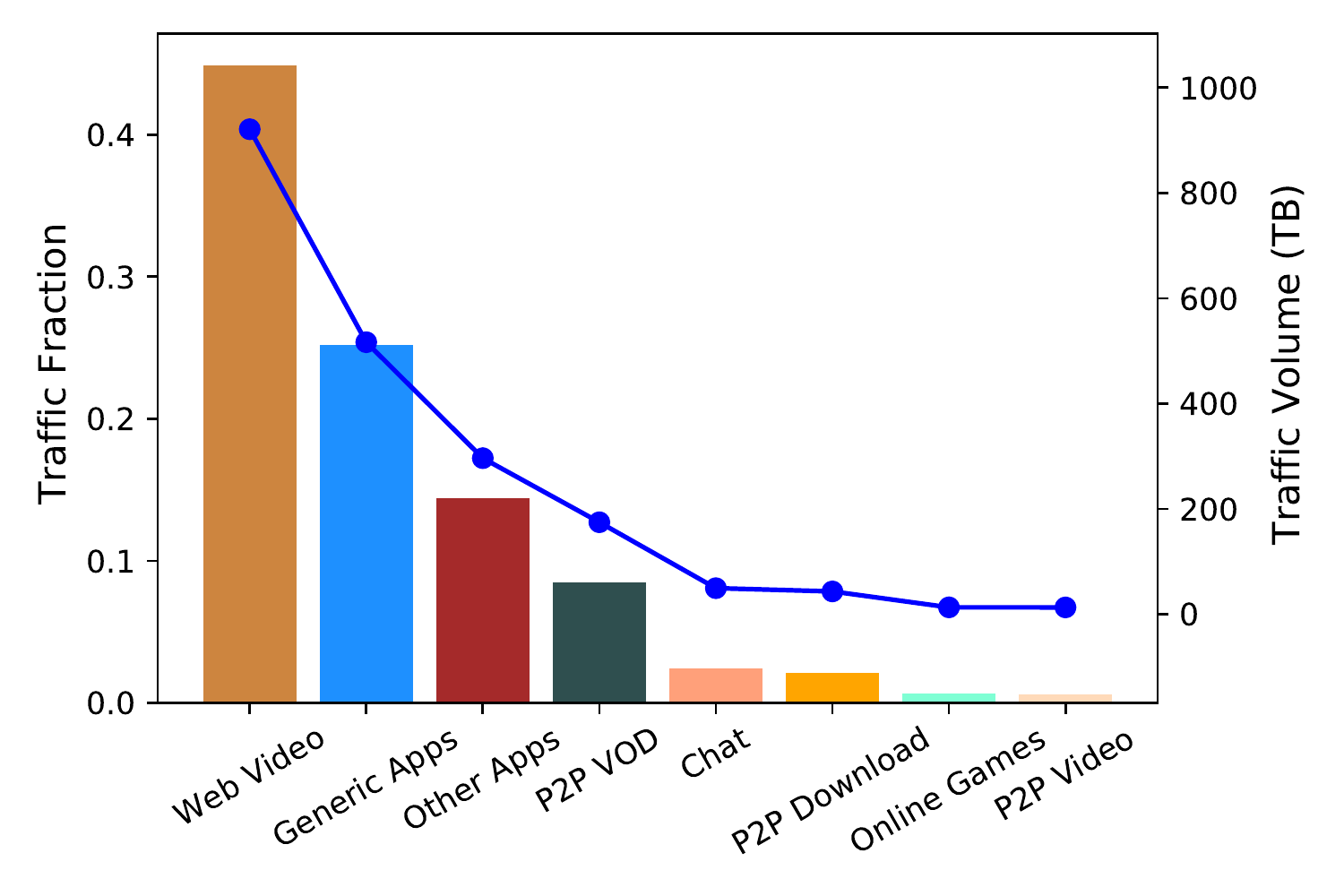}
    \caption{Service traffic volume (line) and fraction of the total (bar) in the city across 8 districts over 31 days. }
    \label{fig:fig3}
\end{figure}

The dataset we employ was collected in a city with over 6 km$^2$ land coverage, administratively divided into 7 districts and 1 core urban area, and with a population of approximately 2.7 million inhabitants. The traffic within each district (D1 to D8) is handled by a dedicated edge node, and the high level structure of the deployment is illustrated in Figure \ref{fig:fig_map}. Traffic data was collected by Deep Packet Inspection (DPI) via port mirroring, between July 1\textsuperscript{st} and 31\textsuperscript{st}, 2020. Traffic was aggregated at session level, with only application type, district identifier, direction (uplink/downlink), total volume and timing information (session start/end) being recorded, to preserved anonymity. In total, 20 service types are distinguished, as summarised in Table \ref{tab:tab_namemap}, where these are sorted in descending order by their volume across the entire deployment, and indexed. The traffic volume distribution for the top-8 services is shown in Figure \ref{fig:fig3}, where bars depict the fraction of the overall volume and the line the corresponding values. Due to the commercially sensitive nature of the dataset, we cannot disclose the precise identity of the city, nor the specific service names for which traffic measurements were collected. 

\begin{table*}[t]
    \setlength{\tabcolsep}{1.5pt}
    \centering
    \begin{tabular}{|l|l|l|l|l|l|l|}
    \hline
\textbf{1:} Web Video & \textbf{2:} Generic Apps  & \textbf{3:} Other Apps&\textbf{4:} P2P VOD & \textbf{5:} Chat & \textbf{6:} P2P Download & \textbf{7:} Online Games  \\ \hline

\textbf{8:} P2P Video  &\textbf{9:} Cloud Storage &\textbf{10:} Shopping Online & \textbf{11:} Live Stream  & \textbf{12:} Music Stream   & \textbf{13:} News Apps   & \textbf{14:} Generic Video  \\ \hline

\textbf{15:} VoIP  &\textbf{16:} Stock Apps& \textbf{17:} Travelling Websites   & \textbf{18:} Mail Apps  &\textbf{19:} Living Apps  &\textbf{20:} Portal Webs & \\ \hline
    \end{tabular}
    \caption{Service names and indexing for the traffic dataset used in our experiments.}
    \label{tab:tab_namemap}
\end{table*}

Over the 31 consecutive days of measurements, we sample the traffic consumption every minute, assuming uniform consumption per session throughout their duration. This is reasonable, given the predominantly short-lived nature of sessions, leading to temporal sequences of 44,640 data points for each service in each region. We normalize service traffic volumes to the 0--1 range, to ensure similar magnitudes during training.
We use an 80/10/10 data split for training, validation, and testing, and train models separately on a region-by-region basis.

\subsection{Benchmarks \& Metrics}

For comparison, we consider the following state-of-the-art DL models as baselines:
\begin{itemize}
    \item \textbf{LSTM}, which is now a classic structure for tackling regression tasks, and has been extensively used for traffic prediction \cite{trinh2018mobile, vinayakumar2017applying}. We implement a three-layer LSTM, which offers an appropriate complexity--effectiveness tradeoff.
    \item \textbf{MTNet}, which was designed for multivariate time series prediction and adopts an encoder-decoder architecture to extract both long- and short-term hidden representations correlation among these~\cite{chang2018memory}. 
    \item \textbf{GraphConv}, which is also aimed at tackling multiple time series predictions \cite{graphconv}, integrating graph convolution \cite{kipf2016semi} and an LSTM network to extract correlations between multiple sequences and temporal patterns. We employ the Spektral library for our implementation \cite{Spektral}.
    \item \textbf{AttentionAR}, a model that we implement based on the Bahdanau Attention structure \cite{bahdanau2014neural} with rolling prediction through a LSTM cell. Attention is used to assign weights to historical input.
\end{itemize}

To evaluate the performance of our proposed models and that of the benchmarks considered, we compute the Mean Absolute Error (MAE) and Root Mean Square Error (RMSE). In essence, these metrics quantify the difference between ground-truth and predicted values, and are defined as follows \cite{krell2020first}:
\begin{equation}\label{eq_mae}
MAE = \frac{1}{S \times F} \sum_{i=1}^{S}\sum_{j=1}^{F} \vert y_{ij} - \hat{y}_{ij} \vert;
\end{equation}

\begin{equation}\label{eq_rmse}
RMSE = \sqrt{ \frac{1}{S\times F}\sum_{i=1}^{S}\sum_{j=1}^{F} (y_{ij} - \hat{y}_{ij})^2 },
\end{equation}
where $S$ is the number of services, $T$ represents the number of prediction steps, and $y_{ij}$ and $\hat{y_{ij}}$ denote the ground truth and respectively predicted traffic volume for service $i$ at timestamp $j$.

\begin{table*}[t]
\centering
\setlength{\tabcolsep}{0.8pt}
\begin{tabular}{c|c|c|c|c|c|c|c|c|c|c|c|c|c|c|c|c}
\toprule
\multirow{2}{*}{Model\textbackslash{District}} &  \multicolumn{2}{c|}{D1}  & \multicolumn{2}{c|}{D2}  & \multicolumn{2}{c|}{D3} & \multicolumn{2}{c|}{D4}  & \multicolumn{2}{c|}{D5} & \multicolumn{2}{c|}{D6} & \multicolumn{2}{c|}{D7} &  \multicolumn{2}{c}{D8}                   \\ \cline{2-17}

& MAE & RMSE & MAE & RMSE & MAE & RMSE & MAE & RMSE & MAE & RMSE & MAE & RMSE & MAE & RMSE & MAE & RMSE \\ \hline
LSTM                        & 45.20     & 125.90       & 47.05 & 136.75          & 51.37   & 145.39          & 21.20    & 60.98        & 21.65   & 63.74        & 15.69    & 44.36        & 18.78      & 56.67    & 50.43    &147.15      \\ 
AttentionAR                 & 61.80    & 176.89       & 63.08 & 184.99         & 65.40      & 181.71      & 24.24   & 66.58         & 27.79     & 81.35        & 19.06   & 53.05          & 23.70      & 70.21     & 64.11     & 180.17      \\ 

GraphConv                   & 107.42   & 300.25         & 110.15  & 327.21        & 116.61    & 317.44       & 36.44   & 99.49         & 40.93   & 123.34         & 23.18     & 66.13    & 34.41     & 108.22        & 118.35         & 340.68    \\ 
MTNet    & 44.15   & 127.37        & 46.78    & 136.54        & 48.97    & 138.71       & 21.86     & 62.33     & 23.70   & 71.22         & 16.40   & 48.22        & 18.54       & 57.21    & 46.51       & 133.26    \\ \hline
\textbf{TMTPN (ours)}               
& \textbf{37.69} & \textbf{106.19}
& \textbf{38.16}  & \textbf{118.76}
& \textbf{43.27}  &\textbf{124.86} 
& \textbf{16.57}  &\textbf{46.80} 
& \textbf{17.99}  & \textbf{53.91}
& \textbf{12.04}  & \textbf{34.94}
& \textbf{14.91}  & \textbf{45.98}
& \textbf{41.49} & \textbf{117.44}\\ \bottomrule
\end{tabular}
\caption{MAE and RMSE performance (in MB) on the multi-service traffic forecasting task with LSTM, AttentionAR, GraphConv, MTNet and our TMTPN across 8 districts.}
\label{tab:tab_mae}
\end{table*}
\subsection{Multi-service Traffic Prediction by TMTPN}

We first examine TMTPN's performance vis-a-vis that of the benchmarks considered, then investigate the influence that input/output lengths have on this. 

\subsubsection{Forecasting Comparison}
We first train and test different models for every district separately, using traffic solely observed within each of these. We take input sequences of length 30 (i.e., 30-min historical data) and predict the traffic volume per service over 5 future timestamps. The obtained results are summarized in Table~\ref{tab:tab_mae}, where lower MAE and RMSE values indicate superior prediction performance. 

Observe that the TMTPN models we propose consistently outperform the state-of-the-art neural networks considered. In particular, when compared with the second best model, MTNet, our TMTPN reduces the MAE and RMSE on avergae by 18.74\% and respectively 18.49\%, across the eight districts. This is because the multi-head attention structure adopted by our design allows the model to jointly extract information from different representation sub-spaces at different points in time, giving higher weights to the most significant historical patterns, to enhance prediction performance. 

The performance of LSTM and MTNet is relatively similar, while AttentionAR largely overestimates traffic demand, indicating that the traditional attention mechanism is not best-suited to multi-service prediction tasks. Finally, GraphConv performs poorly in comparison with our TMTPN and the other two benchmarks we consider. This is likely because no strong spatial relationships between the different services exist in edge network settings. 

\begin{figure*}[t!]
\centering

\subfigure[Service-level predictions in district D1]{\includegraphics[width=\textwidth]{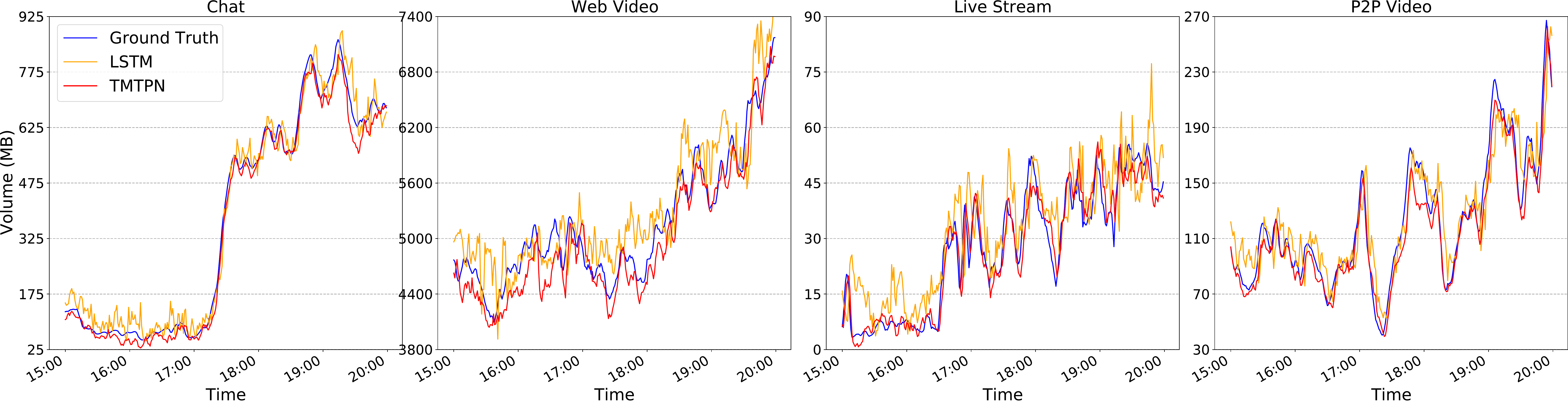}}
\\
\centering
\subfigure[Service-level predictions in district D2]{\includegraphics[width=\textwidth]{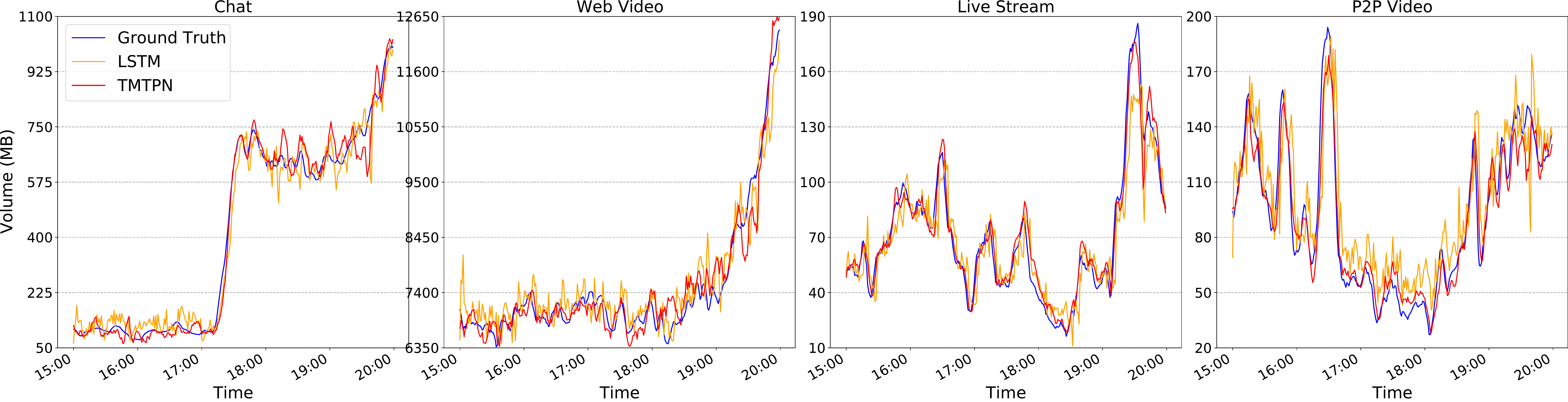}}

\caption{5-step forecasting performance with TMTPN and LSTM at service level over 5 busy hours (15:00 to 20:00 on 29 July, 2020) vs Ground truth.
}
\label{fig:fig_service}
\end{figure*}

To better appreciate the forecasting performance of our proposed TMTPN model at service level, in Table \ref{tab:tab_ser_all} we summarize that across two districts with dissimilar service usage patterns, namely D1 and D2, while in Figure \ref{fig:fig_service} 
we illustrate 5-step prediction instances performed in the two districts across 5-hour windows (busy hours between 15:00 and 20:00 on 29 July, 2020) for 4 randomly selected services (chat, web video, live streaming, P2P video). We only compare TMTPN with LSTM in this and the subsequent experiments, because LSTM appears to be an effective deep learning model that achieves solid performance, which the results in Table \ref{tab:tab_mae} and prior work \cite{he2020graph,zhang2019multi} confirm.
As can been seen from the figure, TMTPN is superior to LSTM, as it tracks more closely the ground truth traffic that would be available under ideal circumstances. This is especially clear to observe on `Chat' traffic forecasting in D1 (sub-figure (a)) and `Live Streaming' traffic in D2 (sub-figure (b)).

\begin{table*}[t]
\centering
\setlength{\tabcolsep}{0.8pt}
\begin{tabular}{c|c|c|c|c|c|c|c|c|c|c|c|c|c|c|c|c|c|c|c|c|c}
\toprule
DIS & Index & 1               & 2               & 3               & 4              & 5              & 6              & 7             & 8              & 9              & 10             & 11            & 12             & 13            & 14            & 15            & 16            & 17            & 18            & 19            & 20            \\ \hline

\multirow{2}{*}{D1} & \textbf{TMTPN}   & \textbf{187.05} & \textbf{250.79} & \textbf{97.91}  & \textbf{52.54} & \textbf{23.71} & \textbf{31.94} & \textbf{5.36} & \textbf{10.85} & \textbf{16.71} & \textbf{24.90} & \textbf{4.32} & 11.01 & \textbf{2.77} & \textbf{1.13} & \textbf{0.59} & \textbf{0.58} & 0.29 & 0.12 & 0.12 & \textbf{0.03} \\ \cline{2-22}
 &LSTM  & 257.40          & 310.18          & 121.12          & 65.28          & 33.13          & 49.63          & 7.92          & 15.47          & 17.39          & 36.40          & 6.26          & \textbf{10.35}          & 2.81          & 2.00          & 0.82          & 0.70          & \textbf{0.26}          & \textbf{0.10}          & \textbf{0.09}          & 0.03          \\ \hline
\multirow{2}{*}{D2} & \textbf{TMTPN}                 & \textbf{270.20} & \textbf{241.52} & \textbf{113.04} & \textbf{56.58} & \textbf{29.85} & \textbf{34.36} & \textbf{6.46} & \textbf{10.15} & \textbf{17.50} & \textbf{8.83}  & \textbf{5.66} & 9.30  & \textbf{2.53} & \textbf{1.04} & \textbf{0.53} & \textbf{0.43} & \textbf{0.08} & 0.07 & 0.13 & \textbf{0.03} \\ \cline{2-22}
&LSTM       & 277.28          & 293.57          & 125.73          & 68.20          & 30.70          & 47.22          & 8.12          & 15.04          & 18.88          & 11.03          & 7.09          & \textbf{8.73}           & 2.65          & 1.49          & 0.67          & 0.45          & 0.08          & \textbf{0.06}          & \textbf{0.12}          & 0.03          \\ \bottomrule
\end{tabular}
\caption{Per-service prediction performance in terms of MAE (MB) in districts D1 and D2. Service names given in Table \ref{tab:tab_namemap} by index.}
\label{tab:tab_ser_all}
\end{table*}

\subsubsection{Impact of Input and Output Length}
\begin{figure}[t]
    \centering
    \includegraphics[width=\linewidth]{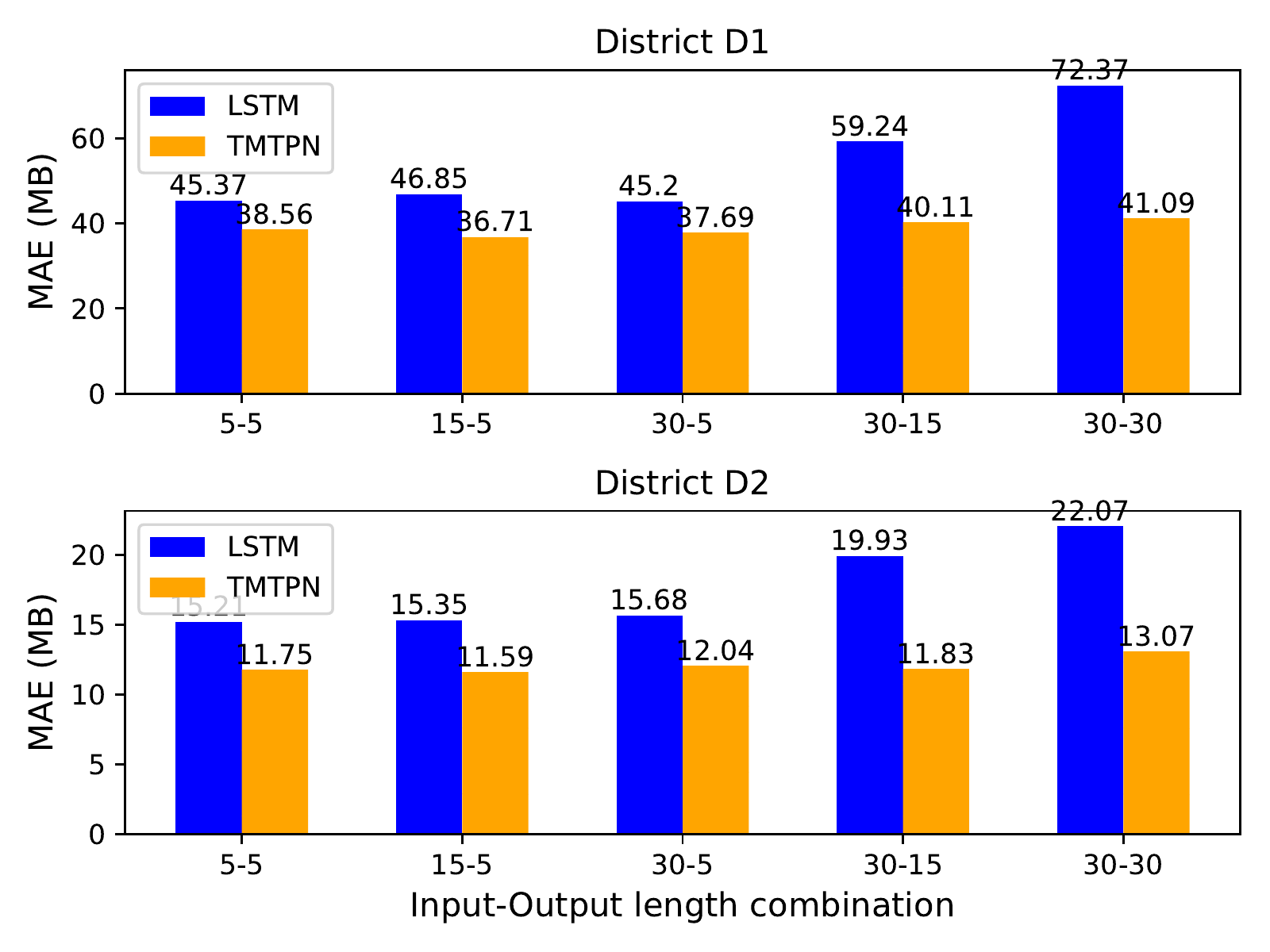}
    \caption{Impact of different Input-Output lengths on forecasting performance (in terms of MAE) with LSTM and our TMTPN in districts D1 (top) and D2 (bottom).}
    \label{fig:fig_len_comp}
\end{figure}

In this subsection, we evaluate the long-term and short-term prediction performance of our TMTPN, focusing again on districts D1 and D2. We examine the MAE across all services as the forecasting horizon varies between 5 and 30 steps, while we also vary the input size, i.e., 5, 15 and 30 historical traffic snapshots. The obtained results are summarized in Figure \ref{fig:fig_len_comp}, where the x-axis represents the combination of input and output length (e.g., 30-5 indicates the model uses the previous 30 minutes traffic data to predict the upcoming 5 minutes traffic demand). The top sub-figure corresponds to district D1 and the bottom to district D2.

Observe that TMTPN is consistently superior to LSTM, as it achieves lower prediction errors. The performance gains grow with the length of the forecasting horizon, with TMTPN reducing the MAE experienced with LSTM on long-term predictions by 43.21\% and 40.77\% in district D1 and D2, respectively. Benefits are also observable short-term, where TMTPN attains 15.02\% and respectively 22.74\% lower MAE than LSTM in the two districts, when the input and output lengths are both 5 (5-5). These gains can be attributed to the multi-head attention mechanism that our design adopts. In addition, the shifted input with look-ahead mask not only enables training parallelization, but also ensures TMTPN can predict the future sequence on a rolling basis, unlike the LSTM, which predicts multiple future steps at once and is thus prone to larger errors. 

Lastly, we note that the input length has only marginal impact on TMTPN's forecasting accuracy, with input size impacting performance slightly differently at the level of the two districts examined. Yet in both cases the best performance is attained with 15 historical snapshots. Based on these results, we argue that if the input length is too short (5), the model may not be able to capture certain periodic information or longer trends. 

\subsection{Multi-service Clustering}
Recall that the aim of service clustering in \name{} is to further improve forecasting performance by grouping services into different clusters, according to their temporal similarity. Here, we demonstrate the benefits of using our WK-means algorithm for this task (hereafter denoted as WASS), as compared to three benchmarks that can be applied to time series data, namely K-means clustering based on Euclidean distance (EUC), K-means based on Cosine similarity \cite{ren2021tree} (COS), and (3) Derivative Dynamic Time Warping (DTW) clustering \cite{he2020graph} implemented with the \texttt{tslearn} library \cite{tslearn}.

\subsubsection{Number of Clusters}
Before comparing the clustering algorithms, the appropriate number of clusters $K$ needs to be determined. The Silhouette score is routinely employed to characterize clustering performance, which is computed as the difference between the mean of the intra-cluster distances and the mean of the nearest-cluster distances, normalized by the maximum between the two \cite{shahapure2020cluster}. The silhouette score is in the $[{-1},1]$ range, with larger values indicating higher quality clustering.

With this, we validate the effectiveness of our WK-means algorithm vis-a-vis that of the benchmarks considered, on the eight districts separately. For each district, $K$ is chosen in the $\{2,\ldots,5\}$ range, and we compute the silhouette score for each $K$ value. The results on district D1 are given in Table \ref{tab:tab_silhouette}, which suggest $K=2$ is the optimum value. The same holds for the vast majority of other districts, with $K=3$ yielding marginally higher silhouette scores (0.01 difference) in 2 out of 32 instances. Hence we select $K = 2$ for all the remaining experiments. 

\begin{table}[t]
\centering
\begin{tabular}{c|c|c|c|c}
\toprule
K\textbackslash{Algorithm} & EUC    & COS    & DTW    & WASS   \\ \midrule
\textbf{2}                 & \textbf{0.8513} & \textbf{0.6891} & \textbf{0.8549} & \textbf{0.5908} \\  \hline
3                          & 0.7879 & 0.6535 & 0.8026 & 0.2638 \\   
4                          & 0.7057 & 0.6182 & 0.6748 & 0.0420 \\    
5                          & 0.6514 & 0.5929 & 0.6742 & 0.0720  \\  \bottomrule
\end{tabular}
\caption{Silhouette score comparison for different numbers of clusters $K$ and the four clustering algorithms considered, in district D1.}
\label{tab:tab_silhouette}
\end{table}

\subsubsection{Clustering Algorithms Comparison}
Next we evaluate forecasting performance with TMTPN when a model is trained individually on services clusters, following grouping of the 20 services into $K=2$ clusters using the proposed WK-means and the benchmark algorithms. We resort again to MAE and RMSE for evaluation and summarize the results obtained in Table \ref{tab:tab_clustering}. 

The results demonstrate that all clustering algorithms can reduce the prediction errors, which is more apparent in districts with larger traffic volumes, such as D1, D2, D3 and D8. Our WASS solution is superior to COS, because cosine similarity gives priority to the direction of two vectors, such as the semantic similarity between two sentences. DTW and EUC are essentially based on the Euclidean distance between traffic magnitude, whereas the ``shape'' of a time series is an important feature when  measuring the similarity between two time series. Our WK-means algorithm (WASS) based on Wasserstein distance possesses such ability, which is reflected in the lower prediction errors obtained (bottom row in Table \ref{tab:tab_clustering}).

\begin{table*}[t]
\centering
\setlength{\tabcolsep}{1pt}
\begin{tabular}{c|c|c|c|c|c|c|c|c|c|c|c|c|c|c|c|c}
\toprule
\multirow{2}{*}{Algorithm\textbackslash{District}} &  \multicolumn{2}{c|}{D1}  & \multicolumn{2}{c|}{D2}  & \multicolumn{2}{c|}{D3} & \multicolumn{2}{c|}{D4}  & \multicolumn{2}{c|}{D5} & \multicolumn{2}{c|}{D6} & \multicolumn{2}{c|}{D7} &  \multicolumn{2}{c}{D8}                   \\ \cline{2-17}

& MAE & RMSE & MAE & RMSE & MAE & RMSE & MAE & RMSE & MAE & RMSE & MAE & RMSE & MAE & RMSE & MAE & RMSE \\ \hline

No Custering & 37.69&106.19   & 38.16&118.76 & 43.26&124.86 & 16.57&46.80 & 17.99&53.91 & 12.04&34.94 & 14.91&45.98& 41.49&117.43 \\ 

EUC   & 31.22&81.26& 31.68&86.91& 33.67&88.83 & 15.00&42.63& 16.49&47.62& 10.80&30.77 & 13.21&39.82  & 33.04&91.57 \\

COS & 35.78&100.92& 37.18&109.03& 41.08&116.10  & 15.67&43.47 & 17.83&52.98& 11.74&34.19  & 14.46&43.59& 38.65&111.33 \\

DTW & 31.22&81.26& 33.35&92.38& 35.92&98.38& 15.45&42.49& 16.49&47.62 & 10.80&30.77& 13.21&39.82 & 34.82&92.63 \\ 

\textbf{WASS(ours)}  & \textbf{28.98}& \textbf{ 81.14}
 &\textbf{30.60} &\textbf{85.09}  & \textbf{32.20}&\textbf{72.87}   & \textbf{14.65}&\textbf{39.36} & \textbf{15.76}&\textbf{46.91}    & \textbf{10.45}&\textbf{30.37}   & \textbf{12.99}&\textbf{39.14}  & \textbf{32.32}&\textbf{87.95}  
\\ \bottomrule
\end{tabular}
\caption{Clustering algorithms comparison based on MAE and RMSE (MB) of forecasts obtained with TMTPN applied on service clusters across the 8 districts.}
\label{tab:tab_clustering}
\end{table*}

\subsubsection{Cluster Visualization} 
\begin{figure*}[ht!]
\centering
\includegraphics[width=\textwidth]{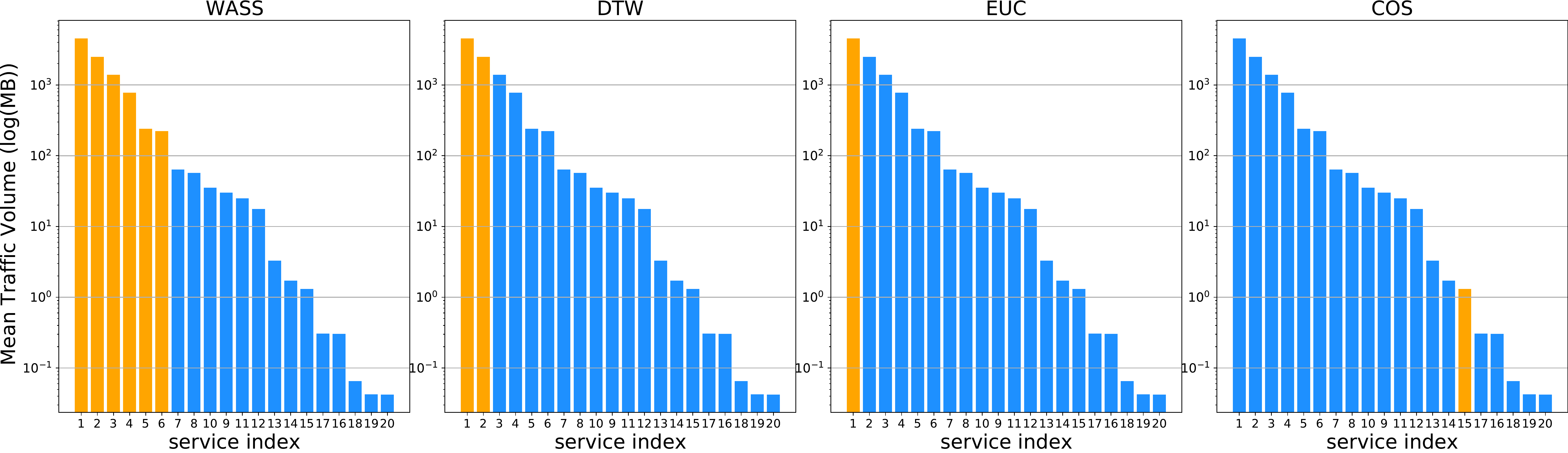}
\caption{Clustering visualization for the four algorithms. Service bars with the same colour are grouped in the same cluster. Service indexes are sorted in descending by traffic volume, and the service name can be obtained by mapping in Table \ref{tab:tab_namemap}}
\label{fig:fig_cluster_comp}
\end{figure*}

To better appreciate where the differences in the performance attained with WK-means and the 3 benchmarks stem from, in Figure \ref{fig:fig_cluster_comp}, we visualize the cluster membership of the different services in a randomly chosen district (D2). DTW and EUC are based on Euclidean distance between each sample and the cluster center, and only services with an extremely large traffic magnitude are categorized into the same cluster. Cosine similarity pays more attention to the difference between two vectors in direction rather than distance or length. In our service clustering task, the traffic magnitude is the primary consideration, therefore cosine similarity is less effective in clustering service time series, which is also confirmed by our previous results reported in Table \ref{tab:tab_clustering}, where COS performs worst than the other three algorithms in all 8 districts.

\subsection{Edge Model Transfer}
Finally, we demonstrate the merits of model transfer in \name{} by showing that reusing models trained at reference nodes within a node cluster, with the aim of reducing computational overhead, does not impact negatively on the forecasting performance. 

\subsubsection{Region Clustering}
Recall that the first step in transferring reference models is to decide the transfer scope. We use K-means clustering to group regions according to the statistical features of all the service traffic time series. We resort again to the silhouette score to determine the optimal number of region clusters, which we computer for $k \in \{2,3,4\}$ in Table \ref{tab:tab_region_score}. We conclude that $k = 2$ produces the highest score and districts D1, D2, D3, D5, and D8 should be grouped together, with the remaining 3 regions belonging to the second edge node cluster.

\begin{table}[ht!]
\centering
\begin{tabular}{c|c|c|c}
\toprule
k & 2    & 3    & 4       \\ \hline
score & \textbf{0.270} & 0.178 & 0.148  \\   \bottomrule

\end{tabular}
\caption{Silhouette score of edge node clustering by K-means with different number of clusters $k$.}
\label{tab:tab_region_score}
\end{table}

\subsubsection{Model Transfer Validation} 

We order the regions according to the overall traffic volume and conclude that D4 and D3 are to be selected as reference nodes for cluster 1 and 2, respectively. We quantify the generalization ability of the models trained by comparing the RMSE when performing forecasting following model transfer (\name{}) versus when models are trained locally at individual region level (Original). To add further perspective and verify our hypothesis that models trained at edge nodes witnessing large traffic volumes have stronger generalization abilities, we also examine the forecasting performance when models are trained on regions where the traffic volume is the lowest among cluster members, prior to transfer (Ctrl-Exp). 

\begin{figure}[ht!]
\centering
\subfigure[Model transfer performance in Edge node cluster 1. The reference models are trained in district D4 (TransMUSE) with the highest traffic volume and D6, which sees the lowest traffic consumption (Ctrl-Exp).  ]{\includegraphics[width=\linewidth]{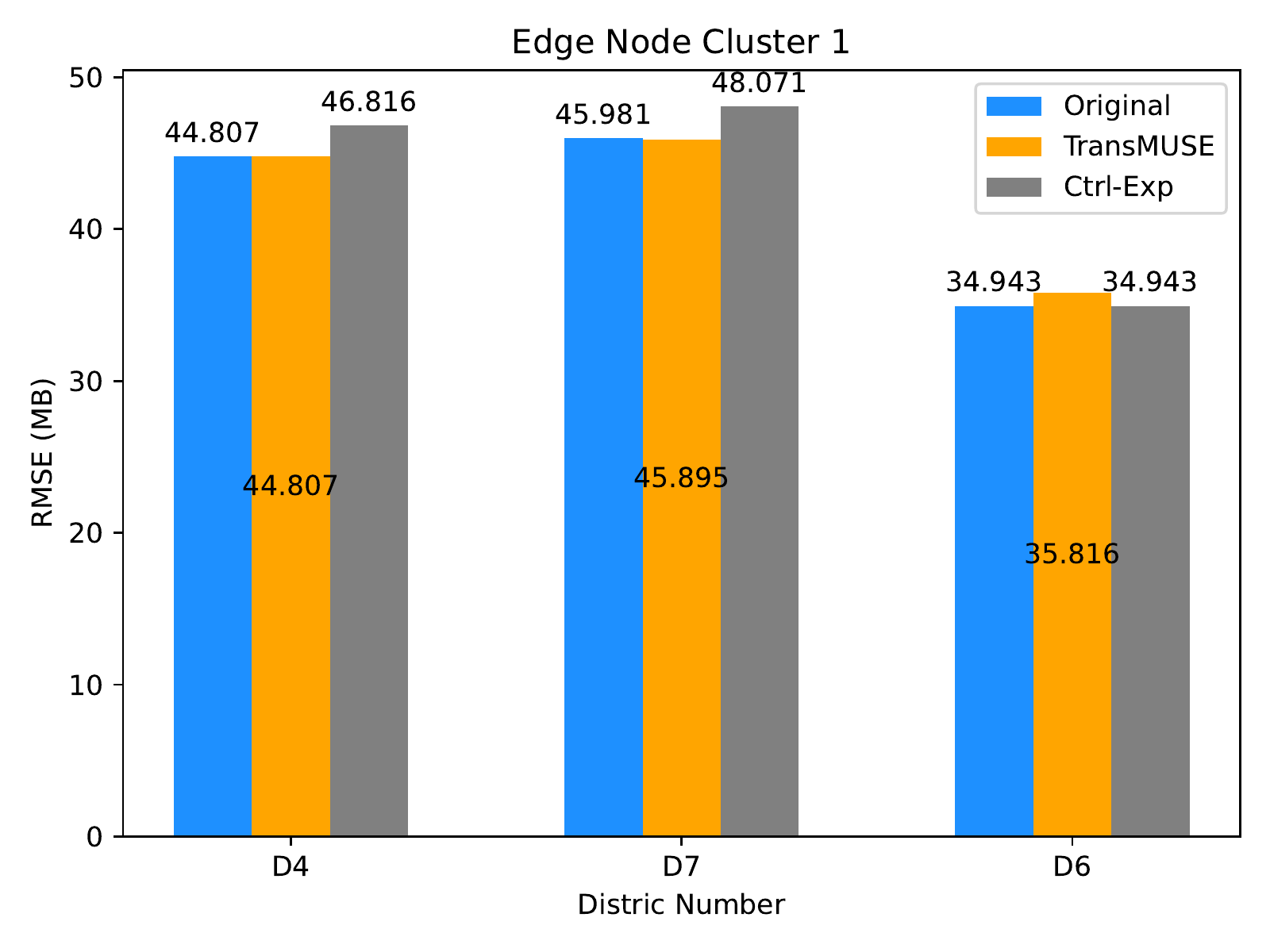}}
\\
\centering
\subfigure[Transfer performance comparison in Edge node cluster 2. The reference models in district D3 (TransMUSE) with the highest traffic volume and D5, which sees the lowest traffic consumption (Ctrl-Exp).
]{\includegraphics[width=\linewidth]{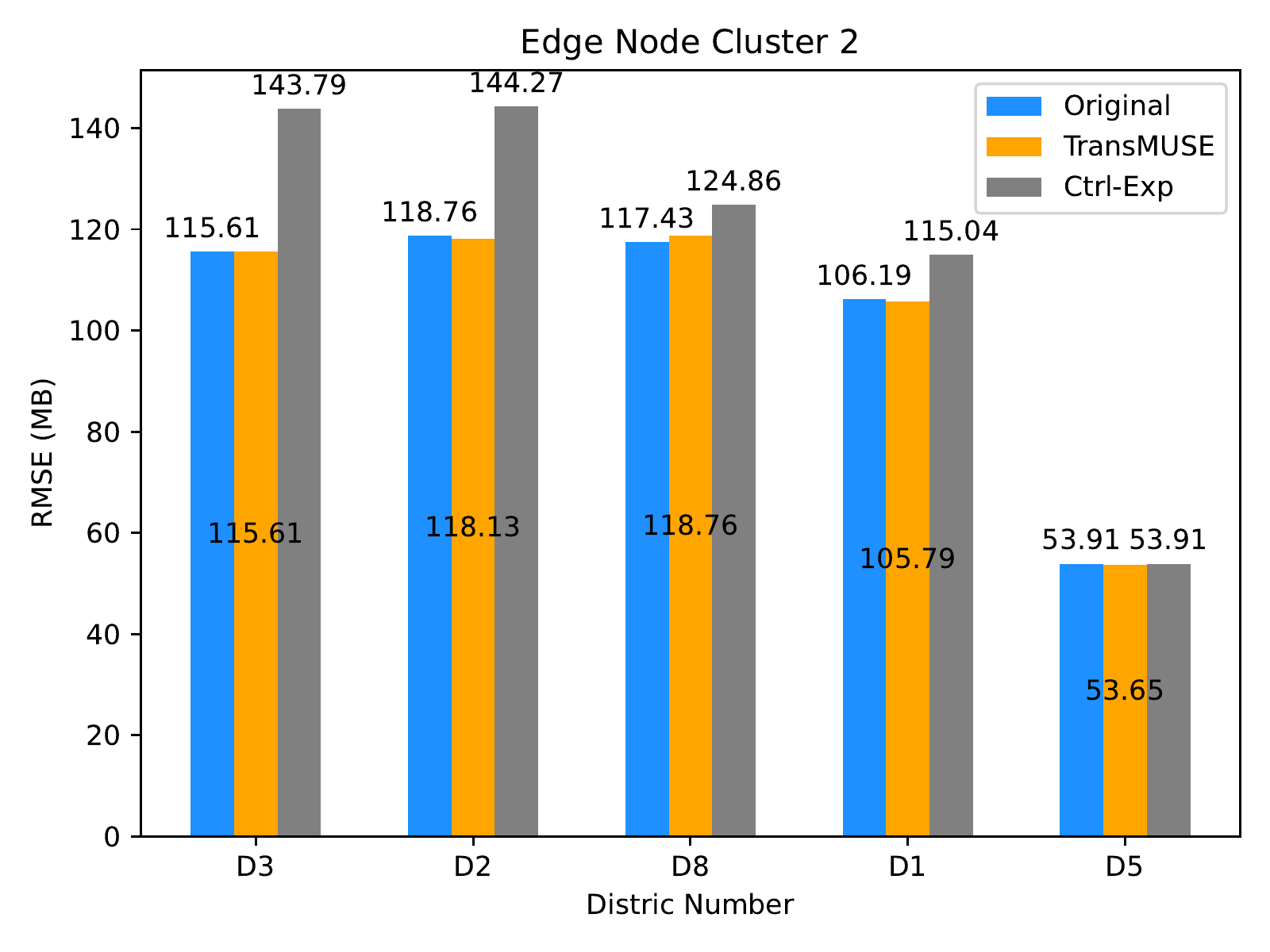}}
\caption{Traffic forecasting performance comparison when model transfer is employed based on highest traffic demand (\name{}), a control experiment where reference models are trained at lightly-loaded edge nodes (Ctrl-Exp), and no edge node clustering is performed, i.e. models trained individually at each location (Original).}
\label{fig:fig_model_transfer}
\end{figure}

The result are illustrated in Figure \ref{fig:fig_model_transfer} for the two clusters, where regions appear in descending order by the overall traffic volume. Observe that when the reference models are trained on regions with the highest traffic demand (\name{}), the RMSE values are almost identical to those obtained when training models individually at each edge node. The largest performance gap is at the level of D6, where a 0.26\% performance degradation is observed in terms of forecasting accuracy (RMSE).
In contrast, if reference models were to be trained at edge nodes with low traffic volumes (D6 in cluster 1 and D5 in cluster 2), the forecasting performance would suffer (Ctrl-Exp). Specifically, the averaged RMSE error over 8 regions is 9.26 MB, which is 92 times larger than with TransMUSE.

We conclude that, as the number of edge nodes increases with the growing adoption of the MEC paradigm, our proposed model transfer strategy will help reduce training time and energy consumption. \name{} will only need to revisit cluster membership and will circumvent the need to persistently collect taffic data in each district.

\section{Related Work}
\label{sec:related-work}
Network traffic prediction is critical to network resource management, optimization and QoS improvement. While this topic has received a lot of attention over the recent years, aspects including service-level traffic forecasting and predicting with low computational overhead have been largely overlooked. Here, we summarise the most relevant work related to our contribution.

\subsection{Traffic Forecasting}
The main approaches to time sequence prediction are State Space Models (SSMs) and sequential models that frequently use deep learning (DL) \cite{wu2020deep}.
The most representative SSMs are Auto-Regressive Integrated Moving Average (ARIMA) models and variants of these, which have been widely adopted for mobile traffic forecasting \cite{adas1997traffic, shu2005wireless, sultan2018call}. Their major drawback is that they require manual parameter selection on a sequence-by-sequence basis. In addition, they perform poorly when inputs exhibit high variability. 

DL has made advances in multiple domains, with Long-Short Term Memory (LSTM) models proven to be superior to traditional models such as ARIMA when predicting wired and wireless traffic \cite{trinh2018mobile,rago2020multi, abbas2017mobile, hua2019deep}. Given that spatial correlations exist between traffic generated at different base stations in wireless networks, LSTM models have been combined with Convolutional Neural Networks (CNNs) to tackle this problem. Zhang et al. proposed a ConvLSTM model to predict multi-service mobile traffic \cite{zhang2019multi} and a graph-sequence spatio-temporal model is introduced in \cite{fang2018mobile} to forecast cellular traffic demand. More recently, attention and transformer architectures demonstrated the ability to handle long sentences in the NLP domain, which subsequently led to their adoption in time series forecasting tasks \cite{he2020graph,wu2020deep}. 

However, none of these prior works builds on the observation that spatial correlations are weak in wired networks and correlations among services matter most.



\subsection{Edge Model Transfer}
As edge computing is getting traction, there have been several research projects focusing on cloud-edge model training based on collaborative learning. He et al. design a collaborative global-local learning scheme that leverages the generalization capability of the global model and the personalization ability of local models to boost the training performance of a graph attention spatio-temporal network (GASTN) for city-wide mobile traffic prediction \cite{he2020graph}. Yan et al. propose COLLA, a collaborative learning framework that allows devices and the cloud to learn collectively user locations \cite{lu2019collaborative}. Zhang et al. design a collaborative cloud-edge computation method for driving behavior modeling, which trains and prunes common models in the cloud and conducts transfer learning at the edge \cite{zhang2019collaborative}. Cartel is proposed in \cite{daga2019cartel} for cloud-edge collaborative learning, aiming at distributing and updating machine learning models across geographically distributed edge clouds.

These works are mostly set on the premise that there exists plenty of data in the cloud to train global models. Edge-edge collaboration, in scenarios where data is largely available only at the network edge, has received less attention. Further, the cost of data transfer overhead has been thus far overlooked, which is non-negligible for network operators.

\section{Conclusion}
\label{sec:conclusions}
In this paper, we tackled network traffic prediction in multi-service edge networks with spatially heterogeneous demands. We proposed \name{}, a framework that groups edge nodes into cohorts and trains transformer-based (TMTPN) models at reference locations, which can be transferred within cohorts without any adaptation. By means of extensive experiments with real-world data, we demonstrated \name{}'s forecasting performance is comparable with that of training individual models with local data at each node. We further propose WK-means, a service clustering routine, which allows to reduce the number of TMTPN models to be maintained for forecasting, based on service similarities. All of these facilitate accurate short- and long-term multi-service traffic prediction with reduced measurement and training costs, which is essential for fine-grained network management. 

\section*{Acknowledgement}
This work was partially supported by the National Natural Science Foundation of China (Grant No.U1909204), the National Key R\&D Program of China (Grant No.2020YFB1-806002), Beijing Natural Science Foundation, China (No. 4202082), the China Scholarship Council and Youth Innovation Promotion Association of Chinese Academy of Sciences (2021168).



\bibliographystyle{elsarticle-num}

\bibliography{TransMUSE}



\end{document}